\documentclass[submission, Proceedings]{SciPost}

\DeclareGraphicsExtensions{.pdf,.png,.jpg}
\graphicspath{{./}}

\begin{document}

\begin{center}{\Large \textbf{
Prospects for $\tau$ lepton physics at Belle II
}}\end{center}

\begin{center}
Michel Hern\'andez Villanueva\textsuperscript{1}
on behalf of the Belle II collaboration
\end{center}

\begin{center}
{\bf 1} Departamento de F\'isica, Centro de Investigaci\'on y de Estudios Avanzados del IPN,
Apdo. Postal 14-740, 07000 Ciudad de M\'exico, M\'exico.
\\
* emhernand@fis.cinvestav.mx
\end{center}

\begin{center}
\today
\end{center}

\definecolor{palegray}{gray}{0.95}
\begin{center}
    \colorbox{palegray}{
        \begin{tabular}{rr}
            \begin{minipage}{0.05\textwidth}
                \includegraphics[width=8mm]{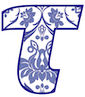}
            \end{minipage}
            &
            \begin{minipage}{0.82\textwidth}
                \begin{center}
                    {\it Proceedings for the 15th International Workshop on Tau Lepton Physics,}\\
                    {\it Amsterdam, The Netherlands, 24-28 September 2018} \\
                    \href{https://scipost.org/SciPostPhysProc.1}{\small \sf scipost.org/SciPostPhysProc.Tau2018}\\
                \end{center}
            \end{minipage}
        \end{tabular}
    }
\end{center}


\section*{Abstract}
{\bf
The Belle II experiment is an upgrade of the Belle detector and will operate at the SuperKEKB energy-asymmetric $e^+e^-$ collider. The accelerator has already successfully completed the first phase of commissioning and first electron positron-collisions in Belle II were observed in April 2018. The design luminosity of SuperKEKB is 8x10$^{35}$ cm$^{-2}$s$^{-1}$ and the Belle II experiment aims to record 50 ab$^{-1}$ of data. Belle II has a broad program of $\tau$ physics, in particular, precision measurements of Standard Model parameters and searches of lepton flavor and lepton number violations, benefiting from the large cross-section of the pairwise $\tau$ lepton production in $e^+e^-$ collisions. In this talk, we will review the $\tau$ lepton physics program of Belle II.
}

\vspace{10pt}
\noindent\rule{\textwidth}{1pt}
\tableofcontents\thispagestyle{fancy}
\noindent\rule{\textwidth}{1pt}
\vspace{10pt}

\section{Introduction}
A B-factory is a machine that collides electrons and positrons at the $\Upsilon(4S)$ resonance energy, producing a large amount of B meson pairs. It provides a clean environment for precision measurements and searches of physics beyond the Standard Model.
The cross section of the process $e^+e^- \to \tau^+\tau^-$ at the $\Upsilon(4S)$ resonance energy is of the same order as the production of a B pair from the $e^+e^-$ collision, then, a B-factory is also a $\tau$ lepton factory.

The first generation of B-factories, BaBar at SLAC and Belle at KEK, have achieved important results in $\tau$ lepton physics, taking advantage of the huge amount of $\tau$ lepton pairs produced in the high luminosity $e^+e^-$ asymmetric collisions at the $\Upsilon(4S)$ resonance energy. Recording a combined sample of 1.5 ab$^{-1}$, corresponding to about 1 billion of tau pair decays, precision measurements of the $\tau$ properties have been performed, such as the mass, lifetime, and branching fractions of leptonic and semileptonic decays. Additionally, limits in electric dipole moment, lepton flavor violation (LFV) and lepton number violation (LNV) decays have been imposed \cite{Bevan:2014iga}. 

\begin{figure}[h!]
    \begin{center}
        \includegraphics[width=13cm]{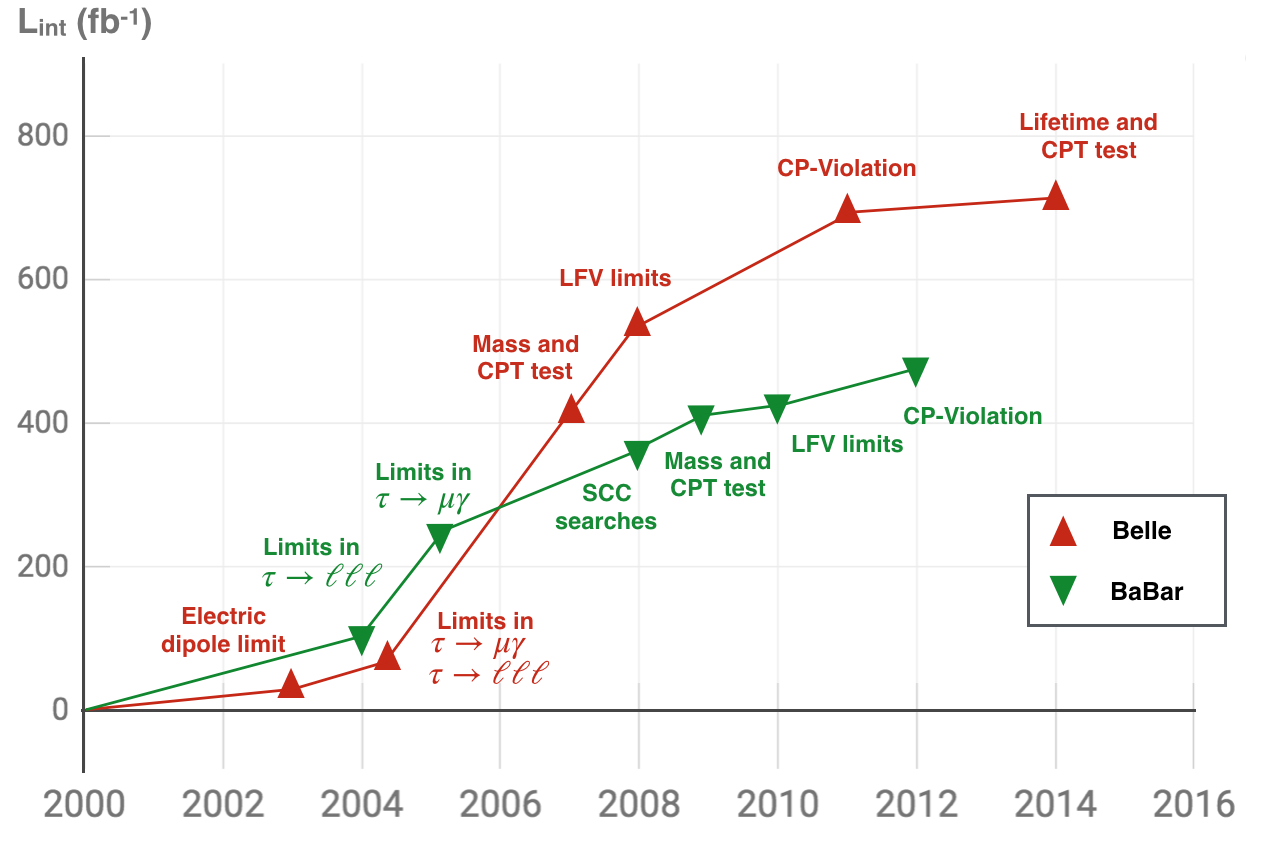} 
        \caption{Highlights of the $\tau$ lepton physics results in the first generation of B-factories, indicating the integrated luminosity accumulated and the year of the measurement.}
        \label{fig:tauHighlights}
    \end{center}
\end{figure}

Figure~\ref{fig:tauHighlights} shows some of the results obtained in the B-factories, indicating the integrated luminosity and the publication year. Most of these results may be improved in the B-factory of next generation, Belle II.

\section{The Belle II experiment}
The Belle II experiment is a detector coupled to the SuperKEKB accelerator, located in Tsukuba, Japan. SuperKEKB is designed to reach 50 nm beam width in the collisions, allowing an instantaneous luminosity of $8.0 \times 10^{35}$ cm$^{-2}$s$^{-1}$.

Belle II keeps the design of the previous detector Belle, with major upgrades in each of their subsystems. The main modifications are:
\begin{itemize}
    \item The vertex detector, which contains two layers of DEPFET pixel (PXD) and four layers of silicon strips (SVD), improving the resolution respect to Belle.
    \item The central drift chamber (CDC) has a larger volume with smaller drift cells.
    \item A completely new particle identification system, using aerogel ring-imaging Cherenkov detectors.
    \item Faster electronics in general. 
\end{itemize}
A complete description of the Belle II detector can be read at reference \cite{abe2010belle}.

From April to July of 2018, Belle II performed the Phase II of commissioning. Detector recorded 500 pb$^{-1}$ of data at $\Upsilon$(4S) energy with the BEAST II detector installed, instead of the vertex detector. BEAST II is used to study the beam background components \cite{lewis2018first}. Next year, in 2019, Belle II will start the Phase III with all the subsystems installed, expecting a full dataset of 50 ab$^{-1}$ by the end of the data taking, in 2025.

\section{First results of $\tau$ lepton physics at Belle II}

\subsection{Reconstruction of $\tau$ pair production}

The reconstruction of tau pair production $e^+e^- \to \tau^+\tau^-$ is performed searching 3-1 prong events in a data sample of 291 pb$^{-1}$. Events in data are required to fire the CDC trigger. Only four charged tracks per event are accepted, with zero net charge and splitting the decay products into two opposite hemispheres by a plane perpendicular to the thrust axis $\hat{\bar{n}}_{thr}$, defined such that
\begin{equation}
V_{thr} = \sum_i \frac{|\vec{p}_i^{cm}\cdot \hat{n}_{thr}|}{\sum \vec{p}^{cm}_i},
\end{equation}
is maximized, with $\vec{p}^{cm}_i$ being the momentum in the center-of-mass system (CMS) of each charged particle and photon. Signal side hemisphere is defined as the one containing a 3-prong decay, while the tag side should contain the 1-prong decay. A pion mass hypothesis is used for all charged tracks, looking for $\tau \to 3\pi \nu$ events in the signal side.

\begin{figure}[h!]
    \begin{center}
        \includegraphics[width=13cm]{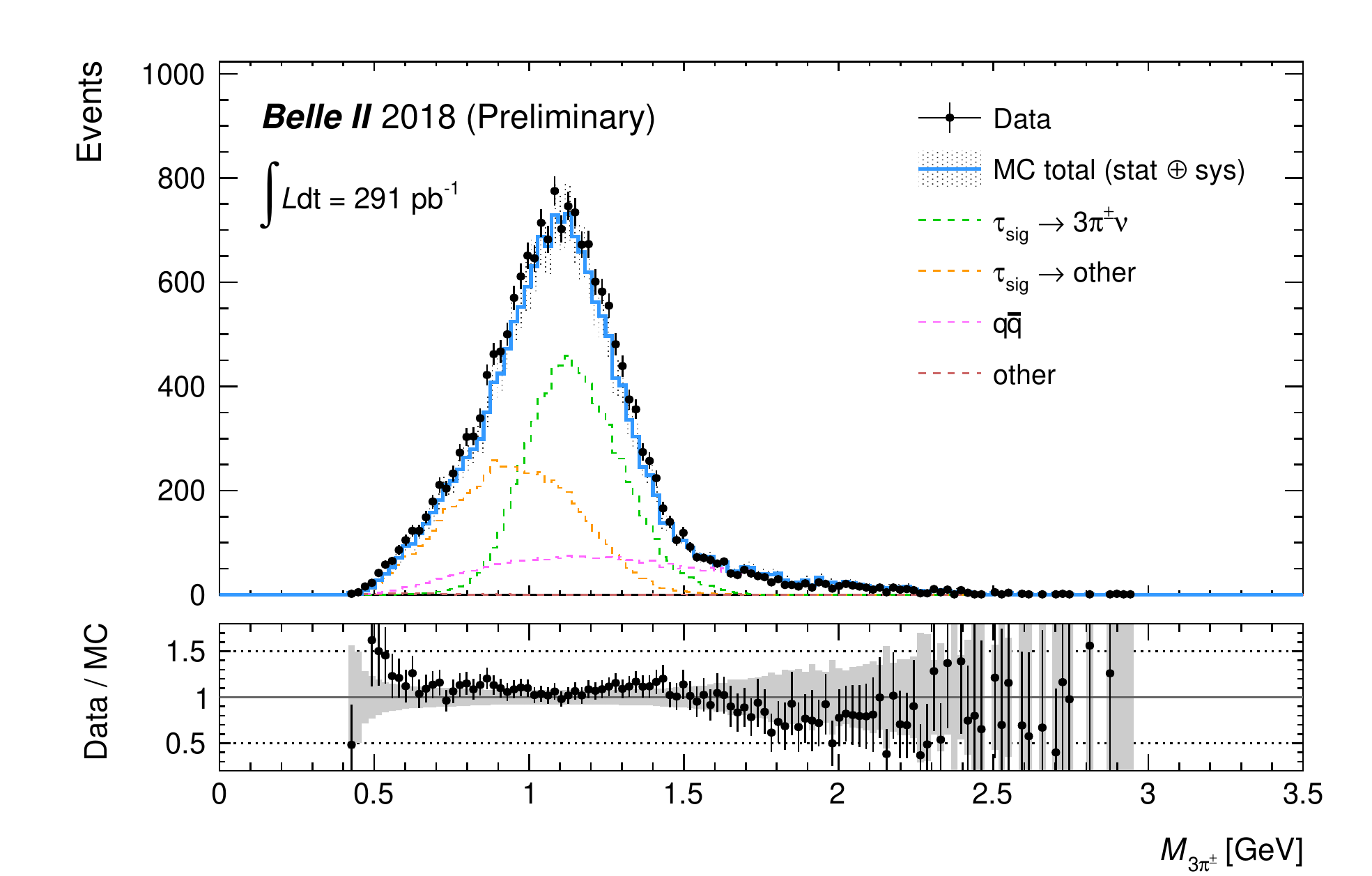} 
        \caption{Invariant mass distribution of the three pions coming from $\tau \to 3\pi \nu$ candidates reconstructed in Phase II data.
            Events in data are required to fire the CDC trigger. MC is rescaled to a luminosity of 291 pb$^{-1}$ and reweighted according to the trigger efficiency measured in data. The error band on the total MC includes the MC statistical uncertainty, the luminosity uncertainty, and the uncertainty associated with the trigger efficiency reweighting.}
        \label{fig:M3pi}
    \end{center}
\end{figure}

\begin{figure}[h!]
    \begin{center}
        \includegraphics[width=12cm]{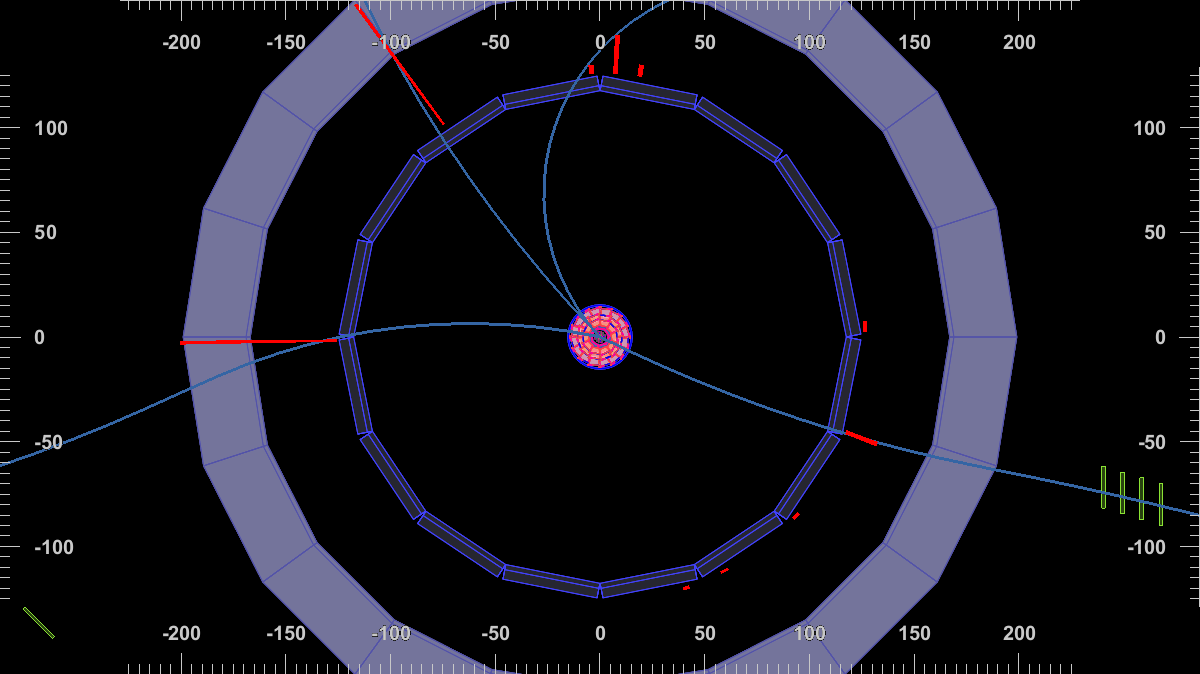} 
        \caption{Event display of the Belle II detector showing a 3-1 prong event, likely a $e^+e^- \to (\tau_{\text{sig}}^{-} \rightarrow 3 \pi \nu$)($\tau_{\text{tag}}^{-} \rightarrow \mu \nu_{\tau} \bar{\nu_\mu}$) candidate reconstructed in Phase II data. }
        \label{fig:eventDisplay}
    \end{center}
\end{figure}

After further selection criteria are applied, 9800 events remain as $\tau$ pair candidates. 
Figure~\ref{fig:M3pi} shows the invariant mass distribution of the three charged pions coming from $\tau \to 3\pi\nu$ candidates, with Monte Carlo (MC) simulated events superimposed.
Figure~\ref{fig:eventDisplay} presents the event display of the Belle II detector with one of the candidates who pass the cuts.

\subsection{$\tau$ lepton mass measurement}

A first $\tau$ lepton mass measurement at Belle II is performed following the method developed by the ARGUS collaboration \cite{albrecht1992measurement}. The pseudomass $M_{min}$, defined by
\begin{equation}
    M_{min} = \sqrt{M_{3\pi}^2 + 2(E_{beam} - E_{3\pi})(E_{3\pi} - P_{3\pi})},
    \label{eq:pseudomass}
\end{equation}
is obtained for each $\tau \to 3\pi\nu$ candidate. Here, $E_{beam}$ is the energy of one of the beams in CMS and $M_{3\pi}$, $E_{3\pi}$, $P_{3\pi}$ represent the invariant mass, the energy and the momentum of the hadronic system of the three pions in CMS, respectively. 

An empirical probability density function (p.d.f.) is used to estimate the $\tau$ lepton mass. The edge p.d.f. used is described by  
\begin{equation}
F(M_{min};a,b,c, m^*) = (a*M_{min} + b)\cdot\arctan\left[(m^* - M_{min})/c \right] + P_1(M_{min})
\label{eq:pdf}
\end{equation}
in which $a,b$ and $c$ are real values and the parameter $m^*$ is an estimator of the $\tau$ lepton mass.

\begin{figure}[h!]
    \begin{center}
        \includegraphics[width=13cm]{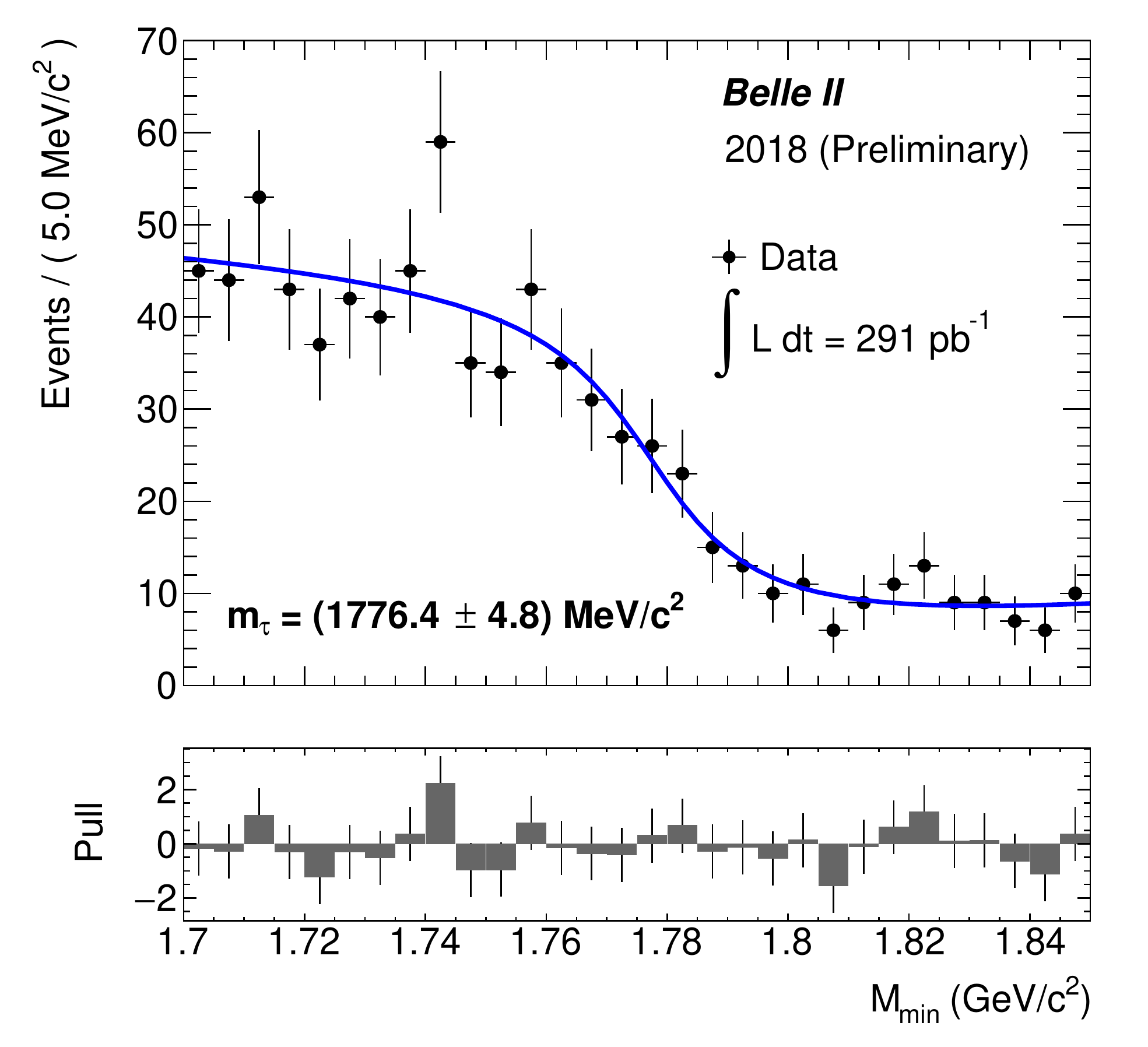} 
        \caption{Distribution of pseudomass $M_{min} = \sqrt{M_{3\pi}^2 + 2(E_{beam} - E_{3\pi})(E_{3\pi} - P_{3\pi})}$ of $\tau_{\text{sig}}^{-} \rightarrow 3 \pi \nu$ candidates reconstructed in Phase II data.  Events are required to fire the CDC trigger. The blue line is the result of an unbinned
            maximum likelihood fit, 
            using an edge function $(a*M_{min} + b)\cdot\arctan\left[(m^* - M_{min})/c \right] + P_1(M_{min})$, in which $m^*$ estimates the $\tau$ lepton mass.
            A mass of $m_\tau = (1776.4 \pm 4.8 $(stat)$)$ MeV/c$^2$ is measured. }
        \label{fig:pseudomass}
    \end{center}
\end{figure}

\begin{figure}[h!]
    \begin{center}
        \includegraphics[width=7cm]{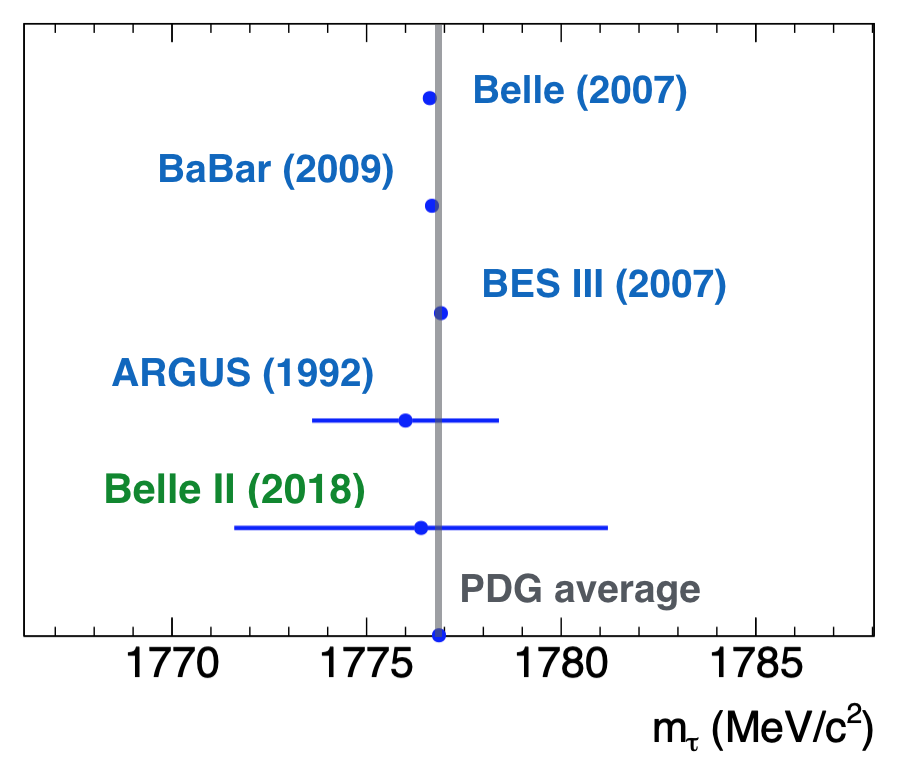} 
        \caption{Comparison between the $\tau$ lepton mass measurement performed by Belle II and the values obtained from previous experiments. $\tau$ lepton mass average reported by the PDG \cite{pdg2018} is also displayed.}
        \label{fig:resultsAll}
    \end{center}
\end{figure}

A fit of the p.d.f. (\ref{eq:pdf}) in the pseudomass region from 1.70 to 1.85 GeV/c$^2$, 
yields a mass measurement of $m_\tau = (1776.4 \pm 4.8 $(stat)$)$ MeV/c$^2$. Figure \ref{fig:pseudomass} shows the pseudomass distribution of the $\tau \to 3\pi \nu$ candidates, with the p.d.f. fitted superimposed. The result is in good agreement with the measurements from previous experiments, as shown in the figure \ref{fig:resultsAll}.

\section{Prospects for $\tau$ lepton physics}

By the end of the data taking, Belle II will have stored 45 billions of $e^+e^- \to \tau^+\tau^-$ events, which will allow the study of $\tau$ physics with high precision measurements. Prospects for $\tau$ lepton physics at Belle II are briefly described. Further details and a more complete description may be found in the Belle II Physics Book \cite{kou2018belle}.

\subsection{Lepton Flavor Violation in $\tau$ decays}
Given the experimental observation of neutrino oscillations, it is known that neutrinos are not massless and lepton flavor is violated. If the Standard Model is extended to include neutrino masses only, the branching ratio of lepton flavor violation (LFV) processes is too small, $\sim 10^{-54}$, to be observed \cite{petcov1976processes, hernandez2018flavor}. The observation of LFV in $\tau$ decays would be a clear indication of physics beyond Standard Model \cite{PhysRevD.89.095014}.

The golden channels for studying charged LFV are $\tau \to 3\mu$ and $\tau \to \mu \gamma$. The first one is a purely leptonic state and the background is suppressed; the second one has the largest LFV branching fraction in models where the decay is induced by one-loop diagrams with heavy particles \cite{Hisano:2002uz, Arganda:2005ji}. 

\begin{figure}[h!]
    \begin{center}
        \includegraphics[width=14cm]{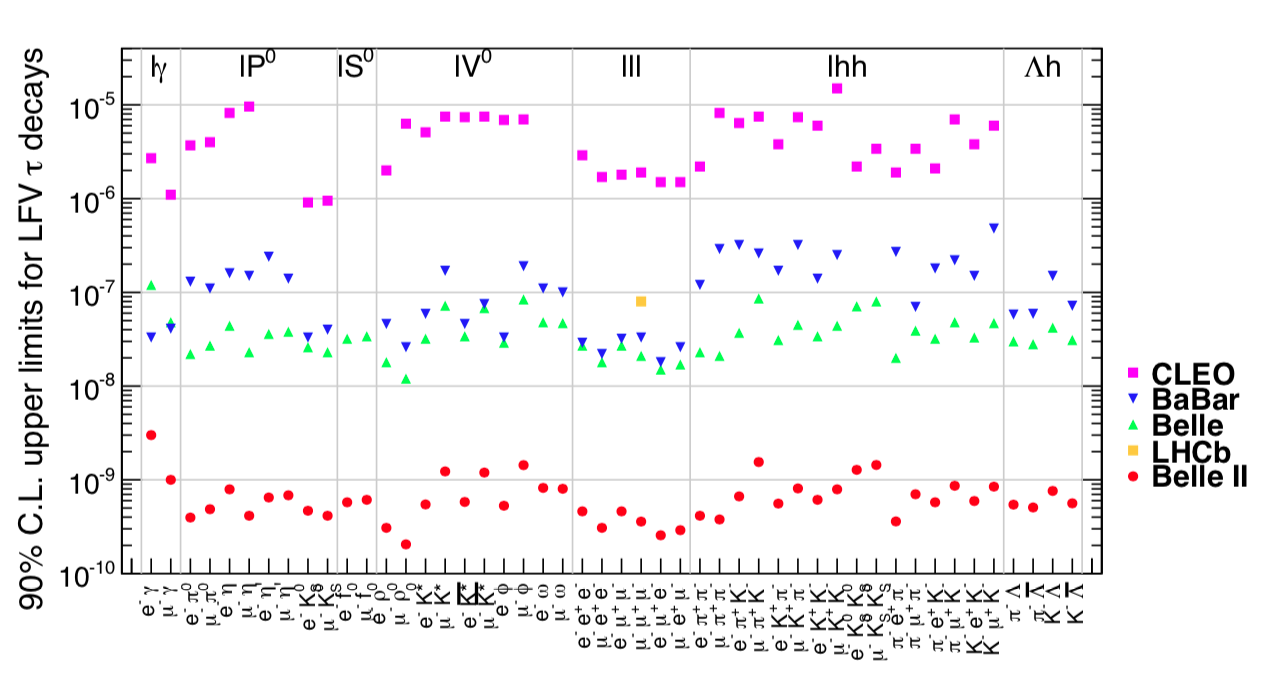} 
        \caption{Current 90\% C.L. upper limits for the branching fraction of LFV $\tau$ decays. Limits imposed by CLEO, BaBar, Belle and LHCb are showed. Additionally, prospects for limits to be imposed by Belle II are indicated with red circles, assuming an integrated luminosity of 50 ab$^{-1}$ \cite{kou2018belle}. The last four decay modes also violates barion number conservation.}
        \label{fig:lfvDecays}
    \end{center}
\end{figure}

Figure \ref{fig:lfvDecays} shows the prospects for upper limits to be imposed in $\tau$ LFV decays according to sensitivity studies described at \cite{kou2018belle} and, for comparison, the limits imposed for previous experiments. With the full dataset expected for the Belle II experiment, 50 ab$^{-1}$, the upper limit for the branching fraction of LFV decays $\tau$ will be reduced by two orders of magnitude.

\subsection{CP violation in $\tau$ decays}
 
The decay of the $\tau$ lepton to final states containing a $K^0_S$ meson will have a nonzero decay-rate asymmetry $A_\tau$, defined by
\begin{equation}
    A_\tau = \frac{\Gamma(\tau^+ \to \pi^+ K^0_S \bar{\nu}_\tau) - \Gamma(\tau^- \to \pi^- K^0_S \nu_\tau)}{\Gamma(\tau^+ \to \pi^+ K^0_S \bar{\nu}_\tau) + \Gamma(\tau^- \to \pi^- K^0_S \nu_\tau)}
\end{equation}
due to CP violation in the kaon sector.  The SM prediction \cite{Grossman:2011zk, bigi2005known} yields 
\begin{equation}
    A^{SM}_\tau = (3.6 \pm 0.1)\times 10^{-3}.
\label{eq:SMtauCP}
\end{equation}
On the experimental side, BaBar is the only experiment that has measured $A_{\tau}$ \cite{lees2012search}, getting
\begin{equation}
    A^{BaBar}_{\tau} = (-3.6 \pm 2.3 \pm 1.1) \times 10^{-3},
\end{equation}
which is 2.8$\sigma$ away from the SM prediction (\ref{eq:SMtauCP}).  An improved measurement of $A_\tau$ is a priority at Belle II.

CP violation could also arise from a charged scalar boson exchange. It can be detected as a difference in the decay angular distributions. Belle searched for CP violation in angular observables of the decay $\tau \to K^0_S \pi \nu$  \cite{PhysRevLett.107.131801}, in which almost all contributions to systematic uncertainty depend on the control sample statistics. So, it is expected that the uncertainties at Belle II will be improved by a factor of $\sqrt{70}$, given the integrated luminosity projected. 

\subsection{Michel parameters}
In the Standard Model, $\tau$ lepton decays due to the interaction with a charged weak current. Leptonic decays such as $\tau \to \ell \bar{\nu_\ell} \nu_\tau$ are of special interest because the absence of strong interactions allows the precise study of electroweak current and its Lorentz structure. 
The most general Lorentz invariant Lagrangian, assuming left-handed neutrinos, is
\begin{equation}
    \mathcal{L} = \frac{4G_F}{\sqrt{2}}\sum_{\substack{N= S,V,T \\ i,j = L,R}} g_{ij}^N \left[ \bar{\Psi_i}(l) \Gamma^N \Psi_n(\nu_l) \right] \left[ \bar{\Psi_m}(\nu_\tau) \Gamma_N \Psi_j(\tau)  \right],
\end{equation}
with $\Gamma^S = 1$, $\Gamma^V = \gamma^\mu$ and $\Gamma^T = \frac{i}{2\sqrt{2}}(\gamma^\mu \gamma^\nu - \gamma^\nu \gamma^\mu)$, being $\gamma^\nu$ the Dirac matrices. Ten non-trivial coupling constants $g^N_{ij}$ describe the interaction. In the SM, with a Lorentz structure V-A of the current, the only non-zero coupling is $g^V_{LL} = 1$.

In the leptonic decay $\tau \to \ell \bar{\nu_\ell} \nu_\tau$, assuming neutrinos are not detected and the spin of the outgoing lepton is unknown, only four bilinear combinations of the coupling constants $g^N_{ij}$ are experimentally accessible. They are called the Michel parameters $\rho$, $\eta$, $\xi$ and $\delta$ \cite{michel1950interaction}. In the SM, $\rho = 3/4$, $\eta = 0$, $\xi = 1$ and $\delta = 3/4$. 

The expected statistical uncertainty of Michel parameters at Belle II is of the order of 10$^{-4}$. Assuming a similar performance as Belle \cite{Epifanov:2017kly}, at Belle II the systematic uncertainties will be the dominant ones. Improvements in the two-track trigger are needed in order to reduce systematic uncertainties during the measurement. 

\subsection{Searches for second class currents in $\tau$ decays}

According to their spin, parity and G-parity (J$^{PG}$), hadronic currents can be classified \cite{weinberg1958charge} as first class currents, with quantum numbers $J^{PG} = 0^{++}, 0^{--}, 1^{+-}, 1^{-+}$; and second class currents (SCC) with $J^{PG} = 0^{+-}, 0^{-+}, 1^{++}, 1^{--}$, being the last ones not discovered yet. 

SCC could be discovered in nuclear processes, with many theoretical and experimental challenges. Another possibility is their study in $\tau$ lepton decays, in which the observation of the decay $\tau^- \to b^-_1 \nu_\tau$ or $\tau^- \to a^-_0 \nu_\tau$ would be a clear signature of SCC \cite {leroy1978tau}.

The most feasible possibility at Belle II is the search of SCC via the decay $\tau \to \eta \pi \nu$, containing or not a intermediate $a_0$ resonance. In the SM, the proposed decay is suppressed by isospin violation through the $\pi^0 - \eta$ mixing, and a branching ratio of Br$(\tau \to \eta \pi \nu)\sim10^{-5}$ is expected \cite{Nussinov:2008gx, Paver:2010mz, Volkov:2012be, Descotes-Genon:2014tla, Escribano:2016ntp}. This decay channel has not been observed yet, given the difficulty in controlling the background. BaBar has set the upper limit Br$(\tau \to \eta \pi \nu) < 9.9 \times 10^{-5}$.

\begin{table}
\begin{center}
\begin{tabular}{|c|c|c|c|}
    \hline 
    BR$_V$ (x10$^5$) & BR$_S$ (x10$^5$) & BR$_{V+S}$ (x10$^5$) & Model \\ 
    \hline 
    0.36 & 1.0 & 1.36 & MDM, 1 resonance \cite{Nussinov:2008gx} \\ 
    \hline 
    [0.2, 0.6] & [0.2, 2.3] & [0.4, 2.9] & MDM, 1 and 2 resonances \cite{Paver:2010mz} \\ 
    \hline 
    0.44 & 0.04 & 0.48 & Nambu-Jona-Lasinio  \cite{Volkov:2012be}\\ 
    \hline 
    0.13 & 0.2 & 0.33 & Analicity, Unitarity \cite{Descotes-Genon:2014tla}\\ 
    \hline 
    0.26 & 1.41 & 1.67 & 3 coupled channels \cite{Escribano:2016ntp} \\ 
    \hline 
\end{tabular} 
\caption{Most recent SM predictions of the branching ratio of the decay $\tau \to \eta \pi \nu_\tau$.}
\label{table:SMpredictions}
\end{center}
\end{table}

\begin{figure}[h!]
    \begin{center}
        \includegraphics[width=14cm]{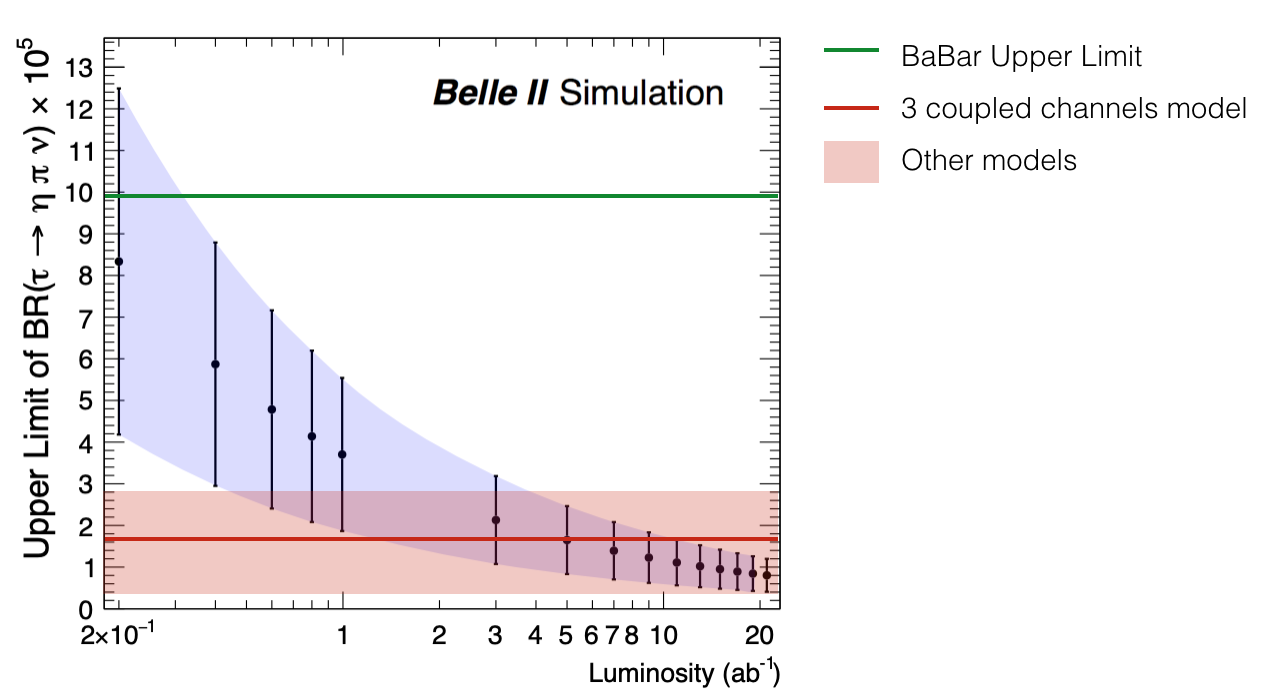} 
        \caption{
        Prospects for 90\% C.L. upper limits to be imposed by Belle II for the branching fraction of the decay $\tau \to \eta \pi \nu$, as function of the integrated luminosity recorded. Red line represents the prediction for branching ratio of the 3 coupled channels model \cite{Escribano:2016ntp} and red band contains other SM predictions described in table \ref{table:SMpredictions}.}
        \label{fig:sccLimits}
    \end{center}
\end{figure}

At Belle II, the large statistics will allow imposing stronger cuts in order to reduce the background. Sensitivity studies performed show Belle II will have full capacity to test models proposed for $\tau \to \eta \pi \nu$ (See table~\ref{table:SMpredictions}). Figure \ref{fig:sccLimits} shows the 90\% C.L. upper limits to be imposed by Belle II for the branching fraction of the decay $\tau \to \eta \pi \nu$. From 1 ab$^{-1}$, Belle II will have the capacity to test SM predictions for the $\tau \to \eta \pi \nu$ decay.
Furthermore, given the suppression expected in the SM, new physics searches are achievable \cite{Garces:2017jpz}.

\section{Conclusions}

Belle II has recorded successfully $\sim$500 pb$^{-1}$ of data during the first collisions performed at SuperKEKB during Phase II. The data has been used mostly for beam background and detector performance studies, showing a healthy operation of all the subsystems. In 2019, full physics program will start and by the end of the experiment, in 2025, Belle II is expected to collect 50 ab$^{-1}$ of data.

The preliminary result on $\tau$ lepton mass measurement obtained from Phase II data, 
$m_\tau = (1776.4 \pm 4.8)$ MeV, is in good agreement with the measurements reported by previous experiments and the average $\tau$ mass value by the PDG. Systematic uncertainties were not considered. Additionally, a good agreement between data and simulations is observed.

The $\tau$ lepton physics program at Belle II will take advantage of the largely integrated luminosity expected, allowing the study of several topics. Limits in branching ratio of LFV decays, CP violation asymmetries and Michel parameters will be improved by two orders of magnitude, but a careful analysis of systematic uncertainties is required. Searches for the $\tau \to \eta \pi \nu$ decays should allow the discovery of SCC and studies of physics beyond the Standard Model.

\bibliographystyle{SciPost_bibstyle} 
\bibliography{ReferencesProceeding.bib}

\begin{thebibliography}{10}
\providecommand{\url}[1]{\texttt{#1}}
\providecommand{\urlprefix}{URL }
\expandafter\ifx\csname urlstyle\endcsname\relax
  \providecommand{\doi}[1]{doi:\discretionary{}{}{}#1}\else
  \providecommand{\doi}{doi:\discretionary{}{}{}\begingroup
  \urlstyle{rm}\Url}\fi
\providecommand{\eprint}[2][]{\url{#2}}

\bibitem{Bevan:2014iga}
A.~J. Bevan \emph{et~al.},
\newblock \emph{{The Physics of the B Factories}},
\newblock Eur. Phys. J. C \textbf{74}(11), 3026 (2014),
\newblock \doi{10.1140/epjc/s10052-014-3026-9}.

\bibitem{abe2010belle}
T.~Abe \emph{et~al.},
\newblock \emph{Belle {II} technical design report},
\newblock arXiv preprint arXiv:1011.0352  (2010).

\bibitem{lewis2018first}
P.~M. Lewis \emph{et~al.},
\newblock \emph{{First Measurements of Beam Backgrounds at SuperKEKB}},
\newblock Nuclear Instruments and Methods in Physics Research Section A:
  Accelerators, Spectrometers, Detectors and Associated Equipment  (2018),
\newblock \doi{10.1016/j.nima.2018.05.071},
\newblock \eprint{1802.01366}.

\bibitem{albrecht1992measurement}
H.~Albrecht \emph{et~al.},
\newblock \emph{{A measurement of the tau mass}},
\newblock Phys. Lett. B \textbf{292}, 221 (1992),
\newblock \doi{10.1016/0370-2693(92)90634-G}.

\bibitem{pdg2018}
M.~Tanabashi \emph{et~al.},
\newblock \emph{{Review of Particle Physics}},
\newblock Phys. Rev. D \textbf{98}(3), 030001 (2018),
\newblock \doi{10.1103/PhysRevD.98.030001}.

\bibitem{kou2018belle}
E.~Kou, P.~Urquijo \emph{et~al.},
\newblock \emph{{The Belle II Physics book}},
\newblock arXiv preprint arXiv:1808.10567  (2018).

\bibitem{petcov1976processes}
S.~T. Petcov,
\newblock \emph{{The Processes $\mu \to e \gamma$, $\mu \to e e \bar{e}$, $\nu'
  \to \nu \gamma$ in the {Weinberg-Salam} Model with Neutrino Mixing}},
\newblock Sov. J. Nucl. Phys. \textbf{25}, 340 (1977),
\newblock [Erratum: Yad. Fiz.25,1336 (1977)].

\bibitem{hernandez2018flavor}
G.~Hern{\'a}ndez-Tom{\'e}, G.~L. Castro and P.~Roig,
\newblock \emph{Flavor violating leptonic decays of $\tau$ and $\mu$ leptons in
  the {S}tandard {M}odel with massive neutrinos},
\newblock arXiv preprint arXiv:1807.06050  (2018).

\bibitem{PhysRevD.89.095014}
A.~Celis, V.~Cirigliano and E.~Passemar,
\newblock \emph{Model-discriminating power of lepton flavor violating
  $\ensuremath{\tau}$ decays},
\newblock Phys. Rev. D \textbf{89}, 095014 (2014),
\newblock \doi{10.1103/PhysRevD.89.095014}.

\bibitem{Hisano:2002uz}
J.~Hisano,
\newblock \emph{{Lepton flavor violating decay of tau lepton in the
  supersymmetric seesaw model}},
\newblock In \emph{{Higher luminosity B factories. Proceedings, 3rd Workshop,
  Shonan Village, Japan, August 6-7, 2002}}, pp. 166--174 (2002),
  \eprint{hep-ph/0209005}.

\bibitem{Arganda:2005ji}
E.~Arganda and M.~J. Herrero,
\newblock \emph{{Testing supersymmetry with lepton flavor violating tau and mu
  decays}},
\newblock Phys. Rev. D \textbf{73}, 055003 (2006),
\newblock \doi{10.1103/PhysRevD.73.055003},
\newblock \eprint{hep-ph/0510405}.

\bibitem{Grossman:2011zk}
Y.~Grossman and Y.~Nir,
\newblock \emph{{CP violation in $\tau \to \nu\pi K_S$ and $D \to \pi K_S$: The
  importance of $K_S-K_L$ interference}},
\newblock JHEP \textbf{04}, 002 (2012),
\newblock \doi{10.1007/JHEP04(2012)002},
\newblock \eprint{1110.3790}.

\bibitem{bigi2005known}
I.~I. Bigi and A.~I. Sanda,
\newblock \emph{{A 'Known' CP asymmetry in tau decays}},
\newblock Phys. Lett. B \textbf{625}, 47 (2005),
\newblock \doi{10.1016/j.physletb.2005.08.033},
\newblock \eprint{hep-ph/0506037}.

\bibitem{lees2012search}
J.~P. Lees \emph{et~al.},
\newblock \emph{{Search for CP Violation in the Decay $\tau^- \to \pi^- K^0_S
  (>= 0 \pi^0) \nu_\tau$}},
\newblock Phys. Rev. D \textbf{85}, 031102 (2012),
\newblock \doi{10.1103/PhysRevD.85.099904, 10.1103/PhysRevD.85.031102},
\newblock [Erratum: Phys. Rev.D 85,099904 (2012)],
\newblock \eprint{1109.1527}.

\bibitem{PhysRevLett.107.131801}
M.~Bischofberger, H.~Hayashii \emph{et~al.},
\newblock \emph{Search for {CP} violation in
  ${\ensuremath{\tau}}^{\ifmmode\pm\else\textpm\fi{}}\ensuremath{\rightarrow}{K}_{S}^{0}{\ensuremath{\pi}}^{\ifmmode\pm\else\textpm\fi{}}{\ensuremath{\nu}}_{\ensuremath{\tau}}$
  decays at {B}elle},
\newblock Phys. Rev. Lett. \textbf{107}, 131801 (2011),
\newblock \doi{10.1103/PhysRevLett.107.131801}.

\bibitem{michel1950interaction}
L.~Michel,
\newblock \emph{{Interaction between four half spin particles and the decay of
  the $\mu$ meson}},
\newblock Proc. Phys. Soc. A \textbf{63}, 514 (1950),
\newblock \doi{10.1088/0370-1298/63/5/311}.

\bibitem{Epifanov:2017kly}
D.~A. Epifanov,
\newblock \emph{{Measurement of Michel parameters in leptonic $\tau$ decays at
  Belle}},
\newblock Nucl. Part. Phys. Proc. \textbf{287-288}, 7 (2017),
\newblock \doi{10.1016/j.nuclphysbps.2017.03.033}.

\bibitem{weinberg1958charge}
S.~Weinberg,
\newblock \emph{Charge symmetry of weak interactions},
\newblock Physical Review \textbf{112}(4), 1375 (1958).

\bibitem{leroy1978tau}
C.~Leroy and J.~Pestieau,
\newblock \emph{$\tau$-decay and second-class currents},
\newblock Physics Letters B \textbf{72}(3), 398 (1978).

\bibitem{Nussinov:2008gx}
S.~Nussinov and A.~Soffer,
\newblock \emph{{Estimate of the branching fraction $\tau \to \eta \pi^-
  \nu_\tau$, the $a_0(980)$, and non-standard weak interactions}},
\newblock Phys. Rev. D \textbf{78}, 033006 (2008),
\newblock \doi{10.1103/PhysRevD.78.033006},
\newblock \eprint{0806.3922}.

\bibitem{Paver:2010mz}
N.~Paver and Riazuddin,
\newblock \emph{{On meson dominance in the `second class' $\tau \to \eta \pi
  \nu_\tau$ decay}},
\newblock Phys. Rev. D \textbf{82}, 057301 (2010),
\newblock \doi{10.1103/PhysRevD.82.057301},
\newblock \eprint{1005.4001}.

\bibitem{Volkov:2012be}
M.~K. Volkov and D.~G. Kostunin,
\newblock \emph{{The decays $\rho^{-}\to\eta\pi^{-}$ and
  $\tau^{-}\to\eta(\eta')\pi^{-}\nu$ in the NJL model}},
\newblock Phys. Rev. D \textbf{86}, 013005 (2012),
\newblock \doi{10.1103/PhysRevD.86.013005},
\newblock \eprint{1205.3329}.

\bibitem{Descotes-Genon:2014tla}
S.~Descotes-Genon and B.~Moussallam,
\newblock \emph{{Analyticity of $\eta \pi $ isospin-violating form factors and
  the $\tau \rightarrow \eta \pi \nu $ second-class decay}},
\newblock Eur. Phys. J. C \textbf{74}, 2946 (2014),
\newblock \doi{10.1140/epjc/s10052-014-2946-8},
\newblock \eprint{1404.0251}.

\bibitem{Escribano:2016ntp}
R.~Escribano, S.~Gonzalez-Solis and P.~Roig,
\newblock \emph{{Predictions on the second-class current decays
  $\tau^{-}\to\pi^{-}\eta^{(\prime)}\nu_{\tau}$}},
\newblock Phys. Rev. D \textbf{94}(3), 034008 (2016),
\newblock \doi{10.1103/PhysRevD.94.034008},
\newblock \eprint{1601.03989}.

\bibitem{Garces:2017jpz}
E.~A. Garcés, M.~Hernández~Villanueva, G.~López~Castro and P.~Roig,
\newblock \emph{{Effective-field theory analysis of the $\tau^- \to
  \eta^{(\prime)} \pi^- \nu_\tau$ decays}},
\newblock JHEP \textbf{12}, 027 (2017),
\newblock \doi{10.1007/JHEP12(2017)027},
\newblock \eprint{1708.07802}.

\end{thebibliography}

\nolinenumbers

\end{document}